# Three-dimensional radiation dosimetry based on optically-stimulated luminescence

M Sadel[1], E M Høye [2], P S Skyt[2], L P Muren[2], J B B Petersen[2] and P Balling[1]
[1]Dept of Physics and Astronomy, Aarhus University, Denmark
[2]Dept of Medical Physic, Aarhus University / Aarhus University Hospital, Denmark

E-mail: msadel@phys.au.dk

**Abstract.** A new approach to three-dimensional (3D) dosimetry based on optically-stimulated luminescence (OSL) is presented. By embedding OSL-active particles into a transparent silicone matrix (PDMS), the well-established dosimetric properties of an OSL material are exploited in a 3D-OSL dosimeter. By investigating prototype dosimeters in standard cuvettes in combination with small test samples for OSL readers, it is shown that a sufficient transparency of the 3D-OSL material can be combined with an OSL response giving an estimated >10.000 detected photons in 1 second per 1mm$^3$ voxel of the dosimeter at a dose of 1 Gy. The dose distribution in the 3D-OSL dosimeters can be directly read out optically without the need for subsequent reconstruction by computational inversion algorithms. The dosimeters carry the advantages known from personal-dosimetry use of OSL: the dose distribution following irradiation can be stored with minimal fading for extended periods of time, and dosimeters are reusable as they can be reset, e.g. by an intense (bleaching) light field.

## 1. Introduction

Modern radiotherapy (RT), such as e.g. intensity-modulated RT or volumetric modulated arc therapy, as well as the emerging techniques using heavy charged particles, e.g. proton therapy, employ complex spatially modulated 3D radiation fields to deliver therapeutic doses during treatment procedures. Thus, quality assurance (QA) before patient treatment is considered as a necessary prerequisite [1]. The basic dosimetric tools in RT are ionization chambers. On the other hand, methods based on luminescent passive detectors such as optically stimulated luminescence (OSL) detectors are also widely applied, especially for personal dosimetry and phantom measurements [2,3]. However, it seems that only 3D dosimeters (consisting of a radiosensitive volume) meet the expectations of complex 3D dose measurements. The currently known materials used for 3D dosimetry are based mostly on polymerizing [4-9] and radiochromic gels or plastics [10-21]. However, the clinical use of these dosimeters has so far been limited, possibly due to different shortcomings of these materials, e.g. in terms of stability, environmental sensitivity or cost of use.

Dosimetry based on OSL relies on the ability of using light for interrogating trapped states in certain crystalline materials. Upon RT, the population in the trapped states carries information on the local dose and when subsequently subjecting the OSL material to light, the trapped electrons are released to the conduction band and recombine with holes to emit photons with an energy exceeding that of the stimulating light. Based on the intensity of this OSL light it is possible to evaluate the absorbed dose [2]. It has previously been demonstrated that two-dimensional (2D) dose distributions can be obtained by imaging the OSL light from a sample during stimulation, e.g. for use in RT [3].

In this paper, a new approach to 3D dosimetry will be presented. We propose a new reusable 3D radiation dosimetry system, based on OSL material embedded homogenously inside a transparent silicone elastomer matrix, a so-called 3D-OSL dosimeter. With an appropriate optical setup, 3D dose distributions can be obtained from the OSL emission.

## 2. Materials and methods

*2.1. The matrix material*
The matrix consists of a transparent polymeric material based on polydimethylsiloxane silicones (PDMS), which is a commercially available silicone consisting of pre-weighed monomer and curing agent. In the current investigation the SYLGARD® 184 Silicone Elastomer Kit from Dow Corning with a curing agent in 1:10 weight ratio was used.

*2.2. The OSL material*
The OSL material used is lithium fluoride (LiF) doped with magnesium, copper and phosphorus (LiF:Mg,Cu,P - MCP). This material is in widespread use as a very well-established and highly sensitive TL detector e.g. in personal dosimetry, but it has also very good OSL properties [3,22-24]. The MCP powder is embedded homogenously inside the silicone matrix which acts as a host of OSL grains.

*2.3. Preparation*
Prototype 3D dosimeters were cast in 1 x 1 x 5 $cm^3$ cuvettes containing 0.3 g of MCP microcrystals mixed homogenously in 4 g of silicone elastomer. The dosimeters were obtained by mixing dry MCP particles into a silicone matrix. In order to check the anticipated OSL signal levels from each 1 $mm^3$ voxel of a 3D dosimeter, standard OSL-reader aluminum trays carrying amounts of OSL material corresponding to those found in 1 $mm^3$ of the 3D dosimeter were prepared. Two concentrations of OSL powder, which were embedded inside the silicone matrix and poured into the trays, were investigated. Sample no. 1 used approximately 0.06 mg of pure OSL powder, while sample no. 2 used 0.2 mg of OSL powder. A reference sample containing clear silicone without powder was also prepared.

*2.4. The read-out system*
The samples were read-out using the standard Risø TL/OSL DA-20 reader. The samples were irradiated with beta radiation to a dose at the level of 1 Gy in all cases. The samples were stimulated with blue light emitting diodes (LEDs), with emission centered at 470 nm and an intensity of ~80 $mW/cm^2$ at the sample. A Hoya UV-340 filter of 7.5 mm thickness and Ø = 45 mm were used as a detection filter. For the analyses of the OSL signal, the period of stimulation was 100 s.

## 3. Results and discussion
Although both the PDMS matrix and the individual OSL particles are transparent, the 3D-OSL dosimeter material appeared slightly misty, see Figure 1. The transparency depended on the amount of MCP powder used, so the concentration of OSL particles must be optimized as a compromise between signal level per volume and overall transparency. Figure 1 shows a prototype fulfilling both requirements, as discussed in more detail below. One of the advantages of using MCP material in combination with the Sylgard 184 silicone elastomer is that the refractive-index match between LiF and Sylgard 184 is quite good for visible wavelengths, which minimizes light scattering from the embedded particles.

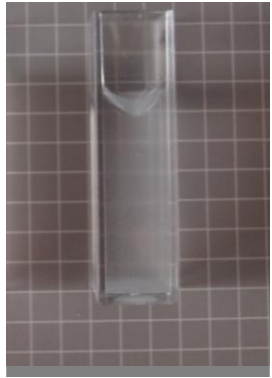

**Figure 1.** A 3D dosimeter in the form of a 1 x 1 x 5 cm$^3$ cuvette containing 4 g of silicone matrix combined with 0.4 g of MCP powder.

During the first second of a standard OSL read-out procedure, ~10,000 and ~40,000 counts were observed from samples 1 and 2, respectively. This corresponded to the anticipated signal levels from each 1mm$^3$ voxels from the 3D dosimeter (at a concentration similar to that of figure 1). We also noted that the silicone matrix in itself – as expected – did not add to the OSL signal (see curve labelled "Empty probe" in Figure 2).

As mentioned, the data presented in the figure 2 were obtained using a commercial OSL reader equipped with UV filters and a PMT-tube. In order to obtain 3D distributions for dosimeters as shown in figure 1, an appropriate optical setup should be applied. For example, by stimulating the OSL dosimeter with a light sheet (e.g. from a laser source), and imaging the luminescence intensity across that sheet (e.g. by a combination of optical filters and a camera), the dose distribution in a plane can be measured. When this plane is shifted across the dosimeter, a 3D dose distribution is directly obtained without the need for inversion algorithms.

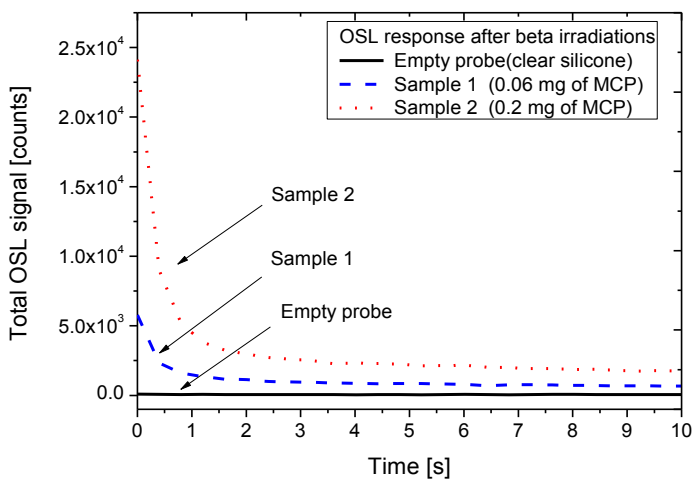

**Figure 2.** OSL decay-curves showing the number of counts in 0.4 s time bins from different 3D-OSL samples obtained by irradiating with a beta source to a dose of 1 Gy. Sample 1 (blue dashed line) contains 0.06 mg of MCP powder, sample 2 (red dotted line) contains 0.2 mg of MCP powder and the last sample contains silicone without MCP powder (black solid line). The OSL was stimulated with a commercial OSL Risø TL/OSL DA-20 reader as described in the text.

An advantage of using the OSL mechanism for dosimetry is that the dosimeters become reusable. It is well known that OSL dosimeters can be reset by temperature or light-bleaching stimulation. The same approach can be used for the OSL material embedded in the silicone host. The new 3D dosimeter also benefits from the well-established dosimetric properties of the MCP material which comprise a wide dynamic range and a linear dose response, together with the potential ability to measure the linear energy transfer (LET) in therapeutic beams. [2, 3, 22-25]. The use of PDMS as a host material makes it simple to cast dosimeters into anthropomorphic shapes and the flexibility of the material even allows for simulating organ deformations during RT in realistic geometries [21].

## 4. Conclusion
We proposed a new reusable 3D dosimeter system based on OSL material embedded homogenously in a transparent silicone matrix and have documented the initial good OSL signal levels for the amount of MCP anticipated in a 1mm$^3$ voxel. The new 3D-OSL dosimeter has the potential to verify complex 3D RT doses with high spatial resolution while maintaining the attractive properties of OSL dosimetry.

## Acknowledgements
This research project was supported by the Danish Cancer Society, award numbers R72-A4486-13-S2 and R56-A2927-12-S2.